\documentclass[12pt,preprint]{aastex}

\slugcomment{Not to appear in Nonlearned J., 45.}

\shorttitle{GRB AFTERGLOWS AS HBLS}
\shortauthors{Wang \& Wei}

\begin{document}


\title{GAMMA-RAY BURST AFTERGLOWS AS ANALOGUES OF HIGH FREQUENCY-PEAKED BL LAC OBJECTS\\}


\author{J. Wang and J. Y. Wei}
\affil{National Astronomical Observatories, Chinese Academy of Science, 20A Datun Road, 
Chaoyang District, Beijing 100012, China}

\email{wj@bao.ac.cn}




\begin{abstract}
The spectral properties from radio to optical bands are compared between the 
18 optically bright Gamma-ray burst afterglows and well established power-spectrum 
sequence in Blazars. The comparison 
shows that the afterglows are well agreement with the well known Blazar sequence
(i.e., the $\nu L_{\nu}(\mathrm{5GHz})$-$\alpha_{\mathrm{RO}}$ correlation,
where $\alpha_{\mathrm{RO}}$ is the broad-band spectral slope from radio to optical bands).
The afterglows are, however, clustered at the low luminosity end of the sequence,
which is typically occupied by high frequency-peaked BL Lac objects. The correlation suggests that 
Gamma-ray burst afterglows share the similar emission process with high frequency-peaked 
BL Lac objects. We further identify a 
deviation at a significance level larger than 2$\sigma$ from the sequence for three 
typical optically ``dark'' bursts. The deviation favors a heavy extinction in optical bands for 
the ``dark'' bursts. The extinction $A_V$ is estimated to be larger than 0.5-0.6 
magnitude from the $\nu L_{\nu}(\mathrm{5GHz})$-$\alpha_{\mathrm{RO}}$ sequence. 
\end{abstract}





\section{INTRODUCTION}

With our current knowledge, Gamma-ray bursts (GRBs) and Blazars (a collective term for 
BL Lac objects and flat-spectrum radio quasars) share the similar observational 
phenomenon involving a relativity jet on the line-of-sight of observers. In addition to the 
observational phenomena,
both classes of the objects share similar central engine: accretion onto a black hole with 
rapid spin, and similar emission processes: synchrotron and possible inverse Compton process (see 
Zhang 2007 for reviews; Urry \& Padovani 1995). In spite of these similarities, 
the two classes of the objects differ from each other in following several respects: central 
engine mass, bulk Lorentz factor and power-spectral energy distribution sequence. 

It is now well known that 
the relativity jets are launched from a supper massive black hole with mass 
$\sim10^{8-9}M_\odot$ for Blazars 
according to their spectroscopic observations (e.g., Lacy et al. 2001). Nevertheless, 
the popular ``collapser'' model indicates that Long-GRBs are produced by the death of 
young massive stars with masses $\geq25M_\odot$ (see Woosley \& Broom 2006 for a review).

Direct measurements of bulk Lorentz factors ($\Gamma$) of jets are available for some Blazars
through the VLBI observations of the superluminoal velocity. The observations reveal a 
typical Lorentz factor $\Gamma\sim10$ for Blazars (e.g., Jorstad et al. 2001; 
Savolainen et al. 2010). However, the GRB fireball model requires 
a larger initial $\Gamma$ ($>100$) to ensure the $\gamma$-ray photons are transparent to 
the $\gamma\gamma\rightarrow e^{\pm}$ process (e.g., Piran 1999). 

The power-spectral sequence has been well established for Blazars by Fossati et al. (1998), 
and was recently confirmed 
by the \it Fermi\rm-detected Blazars (Sambruna et al. 2010), in which the two peaks of 
spectral energy distribution (SED) from radio to $\gamma$-rays shift toward lower frequency for more 
luminous objects. This sequence can be quantitatively described by the tight correlations 
between the photon frequency of the first peak of the SED and the luminosities in different bands.
Several correlations have been found for GRB prompt emission in recent years 
(see Schaefer 2007 for a summary for these correlations). Generally speaking, these 
correlations suggest that GRBs with more powerful prompt emission (or total released energy) tend to 
have higher peaked photon energy in their time averaged $\gamma$-ray spectra 
(e.g., Amati et al. 2009).

In this letter, we compare GRB's radio-optical afterglow emission with the Blazar sequence. 
The study is motivated by the fact that the Lorentz factors $\Gamma$ in GRBs
decrease with both time and radius of the jets as a powerlaw, which means 
the values of $\Gamma$ in the late time 
afterglows could be comparable to Blazars.  
The $\Lambda$ cold dark matter cosmology with parameter $h_0=0.73$, $\Omega_m=0.24$ and $\Omega_\Lambda=0.76$
is assumed throughout our calculations.

\section{SAMPLES}

Frail et al. (2003a) presented a complete catalog of radio afterglows of 
75 pre-\it swift\rm\ GRBs (between 1997 and 2001). The catalog contains the detections 
in the frequencies from 1.4 GHz to 650 GHz for 25 afterglows.
In order to provide a full examination, we further performed
a complete search from the GCN circulars and literature for the radio afterglow detections for the 
bursts triggered later than 2001. The corresponding optical light curves were also compiled from the 
public papers. Our GRB sample finally consists of the radio-optical afterglow detections of 
18 GRBs with measured redshifts. The sample\footnote{Two nearby GRBs, GRB\,980425 and GRB\,060218,
are excluded from the sample to avoid the contamination caused by the radio emission from the 
associated supernovae.} is tabulated in Table 1. For each GRB, Column (1) and 
(2) lists the GRB identification and measured redshift, respectively. 
The time in days at the observer frame since the trigger is 
listed in Column (3). The next two columns present the specific fluxes at 5GHz and at 5500\AA\ 
(V-band) that were observed at the time given in Column (3). The optical V-band flux is 
corrected for the corresponding Galactic absorption, but the local one, 
because the statistical studies indicate that 
the GRB optical afterglows have very small averaged extinction 
around $\langle A_V\rangle\sim0.2$ mag (e.g., Kann et al. 2007).

The last column presents the calculated $k$-corrected monochromatic luminosity at 5\,GHz. The
monochromatic luminosity is derived from the observed flux $f_\nu$ listed in Column (4) as 
$\nu L_\nu(\mathrm{5GHz})=4\pi d_L^2 \nu f_{\nu} (1+z)^{-\alpha-1}$, where $d_L$ is the luminosity distance 
and $\alpha$ the spectral slope defined as $f_\nu\propto\nu^{-\alpha}$. A typical slope 
$\alpha=1/3$ is adopted in the $k$-correction for the slow cooling, optically thin phase (Sari et al. 1998).
The calculated $\nu L_\nu(\mathrm{5\,GHz})$  distribute in a very narrow range roughly from $10^{40}$ to 
$10^{42}\ \mathrm{ergs\ s^{-1}}$, which quality agrees with the similar distribution of the luminosities 
at 8.5\,GHz (Frail 2003).  

Fossati et al. (1998) compiled a sample of Blazars with well sampled SEDs
from radio to $\gamma$-ray. There are totally 110 Blazars with measured redshifts. These objects are 
used as a comparison sample in our subsequent analysis. The authors $k$-corrected all 
the reported specific fluxes to the rest frame according to the empirical spectral slopes. 
Using the $k$-corrected fluxes, we then calculate the corresponding luminosities at 5\,GHz for 
the Blazars by the same method mentioned above.

\section{RESULTS}

Using the calculated average radio-to-$\gamma$-ray SEDs, Fossati et al. (1998) identified a 
Blazar sequence in which there is a systematic trend of the SEDs as a function of the monochromatic luminosities (see also
recently Sambruna et al. 2010).  
This trend results in a tight anti-correlation between the photon frequencies of the first peaks of the SEDs and 
the monochromatic luminosities. However, the frequencies of the first peaks are difficult to be derived for the 
GRB afterglows at the current stage, because the SED modeling requires
multi-wavelength follow-up observations from radio to hard X-rays (even $\gamma$-rays). 
To compare the GRB afterglows with the Blazar sequence, we use the radio-to-optical 
spectral slope $\alpha_{\mathrm{RO}}$ as a proxy of the frequency, both because 
$\alpha_{\mathrm{RO}}$ is correlated with the frequency for the Blazars (e.g., Figure 8 in Fossati et al. 1998) 
and because $\alpha_{\mathrm{RO}}$
can be easily derived for the GRB afterglows. The broad-band spectral slope $\alpha_{RO}$ is estimated by 
following the definition 
\begin{equation}
\alpha_{\mathrm{RO}}=-\frac{\log(f_{\nu_1}/f_{\nu_2})}{\log(\nu_1/\nu_2)}
\end{equation}
for both GRB afterglows and Blazars, where $\nu_1=$5\,GHz, $\lambda_2=c/\nu_2=5500\AA$ ($c$ is the light velocity).
Similar as done in the above section, the spectral slopes $\alpha=-1/3$ and $\alpha=1.2$
is used to carry out $k$-correction in the radio and optical band for the GRB afterglows, 
respectively (Sari et al. 1998).

Figure 1 shows the estimated $\alpha_{\mathrm{RO}}$ plotted against the luminosities at 5\,GHz.
The Blazars from Fossati et al. (1998) are plotted by the red solid circles, and the GRB afterglows from 
Table 1 by the blue open ones. As expected from the Blazar sequence, there is a tight 
correlation between $\alpha_{\mathrm{RO}}$ and $\nu L_\nu$ for the Blazars. A spearman rank-order 
statistical test yields a correlation coefficient $r_s=0.736$ at a significance level $P<10^{-4}$ where 
$P$ is the probability of null correlation. The main conclusion drawn from the figure is that the radio-to-optical 
emission from GRB afterglows closely follow the Blazar sequence. The afterglows are 
clustered at the low luminosity end of the sequence only. The radio luminosities at 5\,GHz of the GRB afterglows 
cover a range from $10^{40}$ to $10^{42}\ \mathrm{erg\ s^{-1}}$, and $\alpha_{\mathrm{RO}}$ from 
0.1 to 0.5. According to the same statistical test, the correlation coefficient 
is slightly enhanced to $r_s=0.798$ with $P<10^{-4}$ when the two classes of the objects are merged into a single 
sample. As an additional illustration, Figure 2 plots the SEDs of the Blazar spectral sequence 
along with that of the afterglow of GRB\,030329 observed at $\sim6$day since the trigger. The 
average Blazar SEDs binned according to the radio luminosity were provided by Fossati et al. (1998) again.
One can see directly from the figure that the SED of GRB\,030329 follows the Blazar sequence at the 
low luminosity end.

By combining the GRB afterglows and Blazars, an unweighted least-square fitting yields a relationship
\begin{equation}
\alpha_{\mathrm{RO}}=(0.27\pm0.02)+(0.09\pm0.01)\log\bigg(\frac{\nu L_\nu(\mathrm{5GHz})}{10^{42}\ \mathrm{erg\ s^{-1}}}\bigg)
\end{equation}
with a 1$\sigma$ dispersion of 0.09.

\section{IMPLICATIONS}

The Blazar sequence was theoretically explained by Ghisellini et al. (1998). Only the very high energy electrons
cool efficiently in a less powerful jet through synchrotron emission, which results in a break in the SED at high 
frequency. In contrast, an external seed photon injection (e.g., broad-line emission from AGNs) causes the powerful 
jet could efficiently cool for low energy electrons through the inverse Compton process. 
According to the unifying scheme,  
the low luminosity end of the $\nu L_{\nu}(\mathrm{5GHz})$-$\alpha_{\mathrm{RO}}$ 
correlation is occupied by the high frequency-peaked BL Lac objects (HBLs).
Without strong external radiation field, the jets in HBLs mainly cool through the synchrotron emission for low energy 
photons and weak SSC for very high energy photons, which predicts a very small Compton dominance, usually 
$\log(L_{\mathrm{c}}/L_{\mathrm{syn}})<0$.

Our study indicates that the GRB radio/optical afterglows follow the Blazars in the 
$\nu L_{\nu}(\mathrm{5GHz})$-$\alpha_{\mathrm{RO}}$ diagram, which implies the afterglows share the 
similar radiation process as that occurring in the HBLs, that is mainly the synchrotron cooling without 
external photons. In fact, the weak inverse Compton components at X-ray band 
have been already identified in a few GRB afterglows with well multi-wavelength detections from radio to X-ray 
(e.g., Harrison et al. 2001; Frail et al. 2003b). The synchrotron model predicates that the peak photon frequency
is $\nu_{\mathrm{peak}}\propto B\Gamma\gamma^2_{\mathrm{peak}}$, where $B$ is the magnetic field, $\Gamma$ the Lorentz 
boost factor, and $\gamma_{\mathrm{peak}}$ the energy of the electrons emitting at the peaks of the SED.
Assuming the recently updated theoretical Blazar sequence $\gamma_{\mathrm{peak}}\propto U^{-1}$ 
($U\propto L/(r^2\theta_j^2)$ is the radiation energy density, where $L$ is the luminosity, and $r$ and $\theta_j$ is 
the distance from center and half opening angel of the jet, respectively) for low luminosity Blazars 
(Ghisellini 2003) is still available for GRBs yields a relationship 
$\nu_{\mathrm{peak}}\propto B\Gamma r^4\theta_j^4/L^2$, which implies $B\Gamma r^4\theta_j^4\sim Br^4\theta_j^3$ 
($\Gamma\sim\theta^{-1}$) might be a constant for GRB late time afterglows taking into account of the tight 
$\nu_{\mathrm{peak}}$-$\nu L_{\nu}$ anti-correlation (see Figure 7 in Fossati et al. 1998).

According to the widely accepted unification scheme for radio-loud sources, the weak Blazars are the 
beamed counterpart of FR I radio galaxies (Urry \& Padovani 1995). The unification
scheme motivate us to suspect the radio emission of GRB afterglow is still detectable even though the 
strongly beamed GRB high energy emission does not point us. These afterglows without high energy prompt 
emission are called as orphan GRB afterglows. In fact, Bower et al. (2007) claimed that they identified 
a single candidate of radio orphan afterglow by examining the radio transient sources observed by VLA.

The nature of optically ``dark'' bursts is still an open issue (e.g., Lamb et al. 2004). 
The dark bursts are now empirically 
defined as the events whose afterglow flux ratio of optical bands to soft X-ray at 3\,keV, 
$\beta_{\mathrm{OX}}$, is less than 
0.5 (Jakobsson et al. 2004). At the present, the debate on the optical darkness mainly focuses on two 
plausible scenarios: the afterglow optical emission is intrinsic faint due to low efficiency (e.g., Jakobsson et al. 2005; 
Urata et al. 2007) or is heavily extincted by the dust (e.g., Zheng et al. 2009; Hashimoto et al. 2010).   

It is emphasized that none of the GRBs listed in Table 1 can be classified as a ``dark'' burst. Table 2 lists
the radio and optical properties of three typical ``dark'' GRBs: GRB\,970828, GRB\,990506, and GRB\,020819. 
For each burst, we list the GRB identification, measured redshift, observing frequency in radio, observation 
time elapsed since the trigger,   
observed flux density in radio, upper limit of R-band flux density, and the calculated monochromatic luminosity
at the corresponding radio frequency. The data are compiled from literature as well.
Along with the $k$-correction, 1) the flux densities (and also $\nu L_\nu$)
at 5\,GHz are transformed from that at 8.5\,GHz by assuming a powerlaw spectral
shape with slope $\alpha=-1/3$ for GRB\,990506 and GRB\,020819; 2)
the upper limits of R-band flux densities are transformed to V-band values by assuming a spectral slope
$\alpha=1.2$ for all the three events.
The three bursts are then over-plotted in Figure 1 by the green open triangles. 
The over-plotted vertical arrows show the calculated spectral slope $\alpha_{\mathrm{RO}}$ 
are underestimated because of the R-band upper limits.

One can clearly see from Figure 1 that all the three ``dark'' bursts significantly deviate 
from the well established correlation (both optically bright GRB afterglows and Blazars) at a significant 
level larger than 2$\sigma$ due to their large $\alpha_{\mathrm{OR}}$. 
The deviation means these bursts
are faint in optical bands with respect to not only X-ray emission, but also radio radiation.

Briefly, there are two approaches that are widely adopted to interpret the optically ``dark'' burst 
puzzle: one is to reduce the value of $\beta_{\mathrm{OX}}$ by enhancing the X-ray emission, 
including long-lived reverse shock for X-ray (e.g., Dai 2004; Genet et al. 2007),
late X-ray prompt emission (Ghisellini et al. 2007), and inverse Compton scattering (Pannaitescu 2008);
an alternative approach is to suppress the optical radiation through heavy dust extinction or intrinsic faintness 
in optical bands. 
The larger $\alpha_{\mathrm{OR}}$ suggests that both X-ray and 
radio emission should be enhanced simultaneously for the optically ``dark'' bursts in the first approach.
In contrast, the faint optical emission in the ``dark'' bursts could be plausibly explained by the 
heavy dust extinction (e.g., Djorgovski et al. 2001), because the radio emission is believed to be 
free of the dust extinction. The extinction scenario was recently supported by Zheng et al. (2009 and references therein)
who found the natural hydrogen column density 
is systematically higher in optically ``dark'' bursts than optically bright events. 
The heavy extinction is also identified in the 
host galaxies of several ``dark'' GRBs (e.g, Levesque et al. 2010).

In the context of the dust extinction scenario, we can roughly estimate the required extinction according to the
$\alpha_{\mathrm{RO}}-\nu L_\nu$ correlation. 
Taking into account of the definition of the slope $\alpha_{\mathrm{RO}}\propto\log(f_{\nu_1}/f_{\nu_2})$, 
we can express the GRB afterglow extinction $A_V$ as a following simple equation $A_V=2.5\Delta\alpha_{\mathrm{RO}}$.
$\Delta\alpha_{\mathrm{RO}}=\alpha_{\mathrm{RO}}-\alpha'_{\mathrm{RO}}$ is the change of the slope due to 
the dust extinction, where $\alpha_{\mathrm{RO}}$ is the observed slope and $\alpha'_{\mathrm{RO}}$
the expected one from the corresponding monochromatic luminosity according to Eq. (2)  
by assuming the $\alpha_{\mathrm{RO}}-\nu L_\nu$ relationship is universal for GRB afterglows.
Note that this approach can only provide us the lower limit of the extinction.
The 1$\sigma$ dispersion of the relationship gives a formal uncertainty of the 
estimated extinction $\delta A_V\sim 0.2$ mag. The required extinctions are estimated for all the three 
``dark'' GRBs and listed in Column (8) in Table 2. Generally speaking, the required
extinctions $A_V>0.5-0.6$ mag, which is obviously larger than the averaged value $\langle A_V\rangle\sim0.2$
obtained from a large sample of GRB optical afterglows (e.g., Kann et al. 2007). 
In fact, a typical dust extinction $A_V\sim1$ mag is identified for some optically ``dark'' GRBs 
(e.g., Perley et al. 2009). Recent near infrared observations indicate that the line-of-sight 
of afterglow of ``dark'' GRB\,080325 shows heavy extinction with an amount $A_V>2$ mag (Hashimoto et al. 2010).

\section{SUMMARY}

We compile a sample of 18 GRB afterglows with multi-wavelength detections from
radio to optical bands. Comparing the afterglows with Blazars shows that the afterglows are well
consistent with the firmly established Blazar sequence 
(i.e., the $\nu L_{\nu}(\mathrm{5GHz})$-$\alpha_{\mathrm{RO}}$
The afterglows are only distributed at the low luminosity end of the sequence, which 
suggests that the GRB afterglows have similar emission mechanism with the 
high frequency-peaked BL Lac objects. The implication on ``dark'' GRBs is also discussed.




\acknowledgments

The authors would like to thank the anonymous referee for his/her comments that improve the 
paper. 
This work was supported by the National Science Foundation 
of China (under grant 10803008) and by the National Basic Research Program of China (grant 2009CB824800).

\clearpage



\begin{figure}
\includegraphics[width = 13cm]{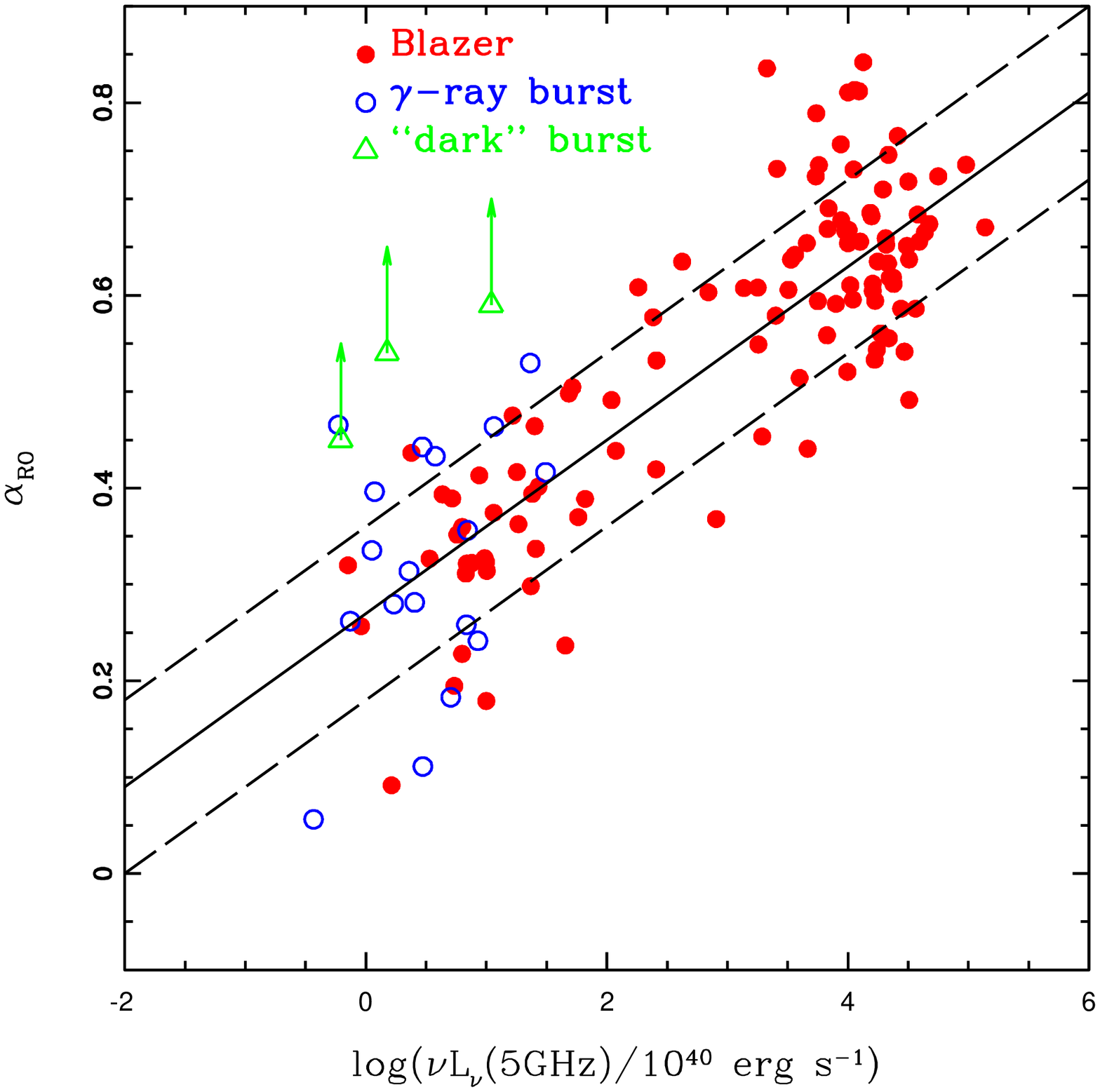}
\caption{Radio-to-optical Broad band spectral slope $\alpha_{\mathrm{RO}}$ plotted 
against the monochromatic luminosity at 5\,GHz at rest-frame. The Blazar sequence is plotted 
by the red solid circles. The 18 optically bright GRBs are over-plotted by the blue open circles, which are 
consistent with the sequence at the low luminosity end. The solid line presents the unweighted least-square fit
to the data when the GRB afterglows and Blazars are merged. The 1$\sigma$ dispersion is marked by the
two dashed lines. In particular, the three green open triangles
and corresponding vertical arrows mark the lower limits for the three typical optically ``dark'' GRBs 
(i.e., GRB\,970828, GRB\,990506, and GRB\,020819). The ``dark'' GRBs deviate 
from the sequence by their large spectral slopes at a significance level $>2\sigma$. 
}
\end{figure}

\begin{figure}
\includegraphics[width = 13cm]{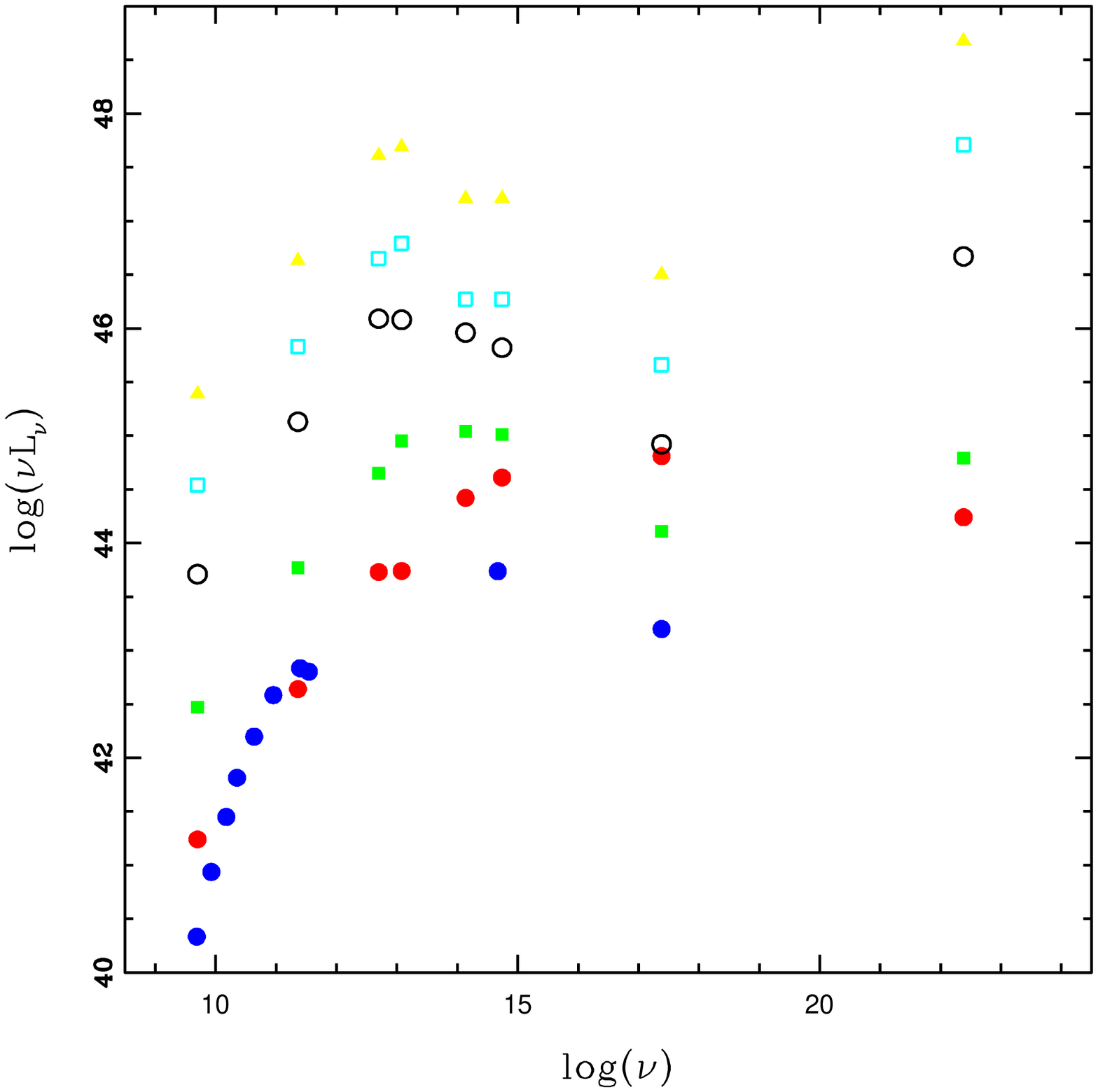}
\caption{
The solid blue circles show the SEDs from radio to soft X-ray of GRB\,030329 afterglow detected at 
$\sim6$ day since the trigger. The multi-wave length data are taken from \it 5GHz, 8GHz, 15GHz, 23GHz, 
43GHz: \rm Berger et al. (2003), \it 90GHz: \rm Kuno et al. 2005, \it 250GHz: \rm Sheth et al. 2003, 
\it 350GHz: \rm Smith et al. 2005, \it V-band: \rm Lipkin et al. 2004, \it 1 keV: \rm Willingale et al. 2004. 
The other symbols plot the average SEDs of the Blazar sequence binned 
according to the radio luminosity, where the SEDs are taken from Fossati et al. (1998).  
}
\end{figure}

\clearpage
\begin{deluxetable}{cccccc}
\tabletypesize{\footnotesize}
\tablecaption{Radio/optical afterglow emission of the 18 optically bright GRBs}
\tablewidth{0pt}
\tablehead{
\colhead{GRB name} & \colhead{z} & \colhead{$\Delta t$} & \colhead{$F_{\mathrm 5GHz}$} &
\colhead{$F_{\mathrm V}$} & \colhead{$\nu L_{\nu}(\mathrm{5GHz})$}\\
  &  & day & $\mu$Jy & $\mu$Jy & $\mathrm{10^{40}\ ergs\ s^{-1}}$ \\
\colhead{(1)} & \colhead{(2)} & \colhead{(3)} & \colhead{(4)} & \colhead{(5)} & \colhead{(6)}\\ 
}
\startdata
970508 &  0.835   &   6.23   &   330  &   5.5   & 5.67  \\ 
980703 &  0.966   &   1.22   &   146  &  10.8   & 3.59  \\
990123 &  1.6     &   0.873  &   104  &  21.5   & 1.41  \\
990510 &  1.619   &   0.72   &   110  &  57.5   & 9.66  \\
991208 &  0.706   &   2.73   &   327  &  28.3   & 3.63  \\
000911 &  1.058   &   4.06   &    60  &   3.7   & 1.82  \\
000926 &  2.066   &   1.18   &    90  &  25.0   &14.48  \\
010222 &  1.477   &   0.26   &    77  &  85.0   & 5.45  \\
010926 &  0.45    &  25.93   &   188  &   1.5   & 0.76  \\
020405 &  0.69    &   3.35   &   160  &   3.6   & 1.68  \\
030329 &  0.168   &   1.44   &  1050  & 693.1   & 0.41  \\
050820 &  2.612   &   2.15   &   256  &  14.6   &73.24  \\
050416A&  0.6535  &   5.6    &   260  &  21.9   & 2.39  \\
070125 &  1.547   &   1.517  &   203  &  51.5   &15.94  \\
071003 &  1.60435 &   2.67   &   256  &   5.1   &22.01  \\
080319B&  0.937   &   2.30   &   204  &   3.3   & 4.58  \\
090323 &  3.57    &   4.99   &   105  &   2.3   &64.01  \\
090902B&  1.882   &   1.31   &   111  &   9.0   &14.18  \\
\enddata
\end{deluxetable}


\begin{deluxetable}{ccccccccc}
\tabletypesize{\footnotesize}
\tablecaption{Radio/optical afterglow emission of the three typical optically ``dark'' GRBs}
\tablewidth{0pt}
\tablehead{
\colhead{GRB name} & \colhead{z} & \colhead{$\Delta t$} &  $\nu$ & \colhead{$F_\nu$} &
\colhead{$F_{\mathrm R}$} & \colhead{$\nu L_{\nu}$} & \colhead{$A_V$} & \colhead{Reference}\\
  &  & days & GHz & $\mu$Jy & $\mu$Jy & $\mathrm{10^{40}\ ergs\ s^{-1}}$ & mag & \\
\colhead{(1)} & \colhead{(2)} & \colhead{(3)} & \colhead{(4)} & \colhead{(5)} & \colhead{(6)} & \colhead{(7)} & \colhead{(8)} & \colhead{(9)}\\
}
\startdata
970828 &  0.9578  &   5.6    &  4.86  &   99  &   $<0.5$  & 1.5   & $>0.63$ & Groot et al. 1998\\
990506 &  1.3     &   1.66   &  8.46  &  447  &   $<1.2$  & 1.3   & $>0.58$ & Taylor et al. 2000\\
020819 &  0.41    &   1.5    &  8.46  &  315  &   $<4.4$  & 0.74  & $>0.50$ & Jakobsson et al. 2005\\
\enddata
\end{deluxetable}



\clearpage




\end{document}